\documentclass[epj]{svjour}
\usepackage{graphicx,latexsym,amssymb,amsmath,color,multirow,mathrsfs,booktabs,ifpdf}
\usepackage{dcolumn}
\usepackage{bm}

\usepackage{float, varioref}

\begin{document}
\title{Equation of state of asymmetric nuclear matter and 
	 the tidal deformability of neutron star}
\author{Ngo Hai Tan\inst{1,2}\and Dao T. Khoa\inst{3} \and Doan Thi Loan\inst{3}}
\institute{Faculty of Fundamental Sciences, Phenikaa University, Hanoi 12116, Vietnam \and  
Phenikaa Institute for Advanced Study (PIAS), Phenikaa University, Hanoi 12116, Vietnam
\and Institute for Nuclear Science and Technology, VINATOM, Hanoi 100000, Vietnam}
\date{Received: date / Revised version: date}
\abstract{Neutron star (NS) is  a unique astronomical compact object where the 
four fundamental interactions have been revealed from the observation and studied 
in different ways. While the macroscopic properties of NS like mass and radius 
can be determined within the General Relativity using a realistic equation 
of state (EOS) of NS matter, such an EOS is usually generated by a nuclear structure 
model like, e.g., the nuclear mean-field approach to asymmetric nuclear matter. 
Given the radius of NS extended to above 10 km and its mass up to twice the solar 
mass, NS is expected to be tidally deformed when it is embedded in a strong tidal 
field. Such a tidal effect was confirmed unambiguously in the gravitation wave 
signals detected recently by the LIGO and Virgo laser interferometers from 
GW170817, the first ever direct observation of a binary NS merger.       
A nonrelativistic mean-field study is carried out in the present work within
the Hartree-Fock formalism to construct the EOS of NS matter, which is then used 
to determine the tidal deformability, gravitational mass, and radius of NS. 
The mean-field results are compared with the constraints imposed for these 
quantities by the global analysis of the observed GW170817 data, and a strong 
impact by the incompressibility of nuclear matter on the hydrostatic 
configuration of NS is shown.} 
\PACS{       
     {21.65.+f}{Nuclear matter} \and
     {26.60.+c}{Nuclear matter aspects of neutron stars} \and
     {04.30.Tv}{Gravitational-wave astrophysics}   
		} 
\maketitle

\section{Introduction}
 \label{sec1}
Neutron star (NS) is an astronomical compact object consisting of the lepton and 
baryon matter under the extreme conditions, and it has been therefore a long standing 
research object of the modern nuclear physics and nuclear astrophysics. The equation 
of state (EOS) of high-density nuclear matter (NM) in the NS core is the main synergy 
between nuclear physics and astrophysics. Using a realistic EOS of asymmetric NM 
characterized by a large neutron excess, the General Relativity not only explains 
the compact shape of NS in the hydrostatic equilibrium with its maximum mass limited 
to about twice the solar mass, but also predicts the interesting behavior of NS in 
the presence of a strong gravitational field. In particular, the shape of NS is tidally 
deformed to gain a nonzero quadrupole moment under the gravitational field formed 
by the mutual attraction of two coalescing neutron stars \cite{Hinderer08,Hinderer10}. 
The gravitational wave (GW) emitting from such a binary NS merger has been expected 
to be detectable by the GW observatories on Earth. Although some indirect evidences 
of the NS merger were suggested earlier, the first direct observation of a binary 
NS merger has been reported on August 17, 2017, with the GW signals detected by LIGO 
and Virgo laser interferometers \cite{Abbott17}, and the $\gamma$ ray burst detected 
by the Fermi $\gamma$ ray space telescope \cite{Abbott17c}. This exciting event is 
now widely referred to as GW170817, which opened a new era of the multimessenger astronomy. 

The GW and $\gamma$ ray burst emitted from GW170817, as well as the x ray, infrared 
radiations, and visible light detected later by more than 70 telescopes provide 
a huge database for the nuclear physics and astrophysics research. By analyzing 
the observed GW signals from GW170817 \cite{Abbott18}, the tidal deformability 
of NS has been determined and translated into a constraint for the gravitational 
mass and radius of NS. This result soon became an important reference for numerous 
nuclear physics studies of NS, and further constraints for the EOS of NM and 
hydrostatic configuration of NS have been suggested 
\cite{Fattoyev18,Annala18,De18,Malik18,Dietrich20}.

A realistic EOS of NS matter (the dependence of the pressure of NS matter on the 
mass-energy density) is the key input for the determination of the macroscopic 
properties of NS like gravitational mass, radius, inertia moment and tidal deformability. 
For this purpose, a nonrelativistic mean-field study of the EOS of NM within the 
Hartree-Fock (HF) formalism has been done in the present work, using several 
versions of the (in-medium) density dependent nucleon-nucleon (NN) interaction 
\cite{Loan11,Tan16}. Using the EOS of asymmetric NM obtained from the HF calculation, 
the np$e\mu$ matter of the uniform NS core is generated from the charge neutrality 
and lepton number conservation implied by the $\beta$ equilibrium. The EOS of the 
np$e\mu$ matter is then used to determine the tidal deformability, gravitational mass
and radius of NS, and a detailed comparison with the constraints imposed for these 
quantities by the GW170817 observation is made.   
 
\section{Hartree-Fock formalism for the EOS of nuclear matter}
 \label{sec2}
The nonrelativistic HF method \cite{Loan11} is used in the present work to model
the EOS of asymmetric NM at zero temperature, which is characterized by the 
neutron and proton number densities, $n_{\rm n}$ and $n_{\rm p}$, or equivalently 
by the total baryon number density $n_{\rm b}=n_{\rm n}+n_{\rm p}$ and neutron-proton 
asymmetry $\delta=(n_{\rm n}-n_{\rm p})/n_{\rm b}$. The total HF energy density of NM 
is obtained as 
\begin{align}
\mathcal{E}=\mathcal{E}_{\rm kin}+{\frac{1}{ 2}}\sum_{k \sigma \tau}
\sum_{k'\sigma '\tau '} [\langle{\bm k}\sigma \tau, {\bm k}' \sigma' \tau'
 |v_{\rm d}|{\bm k}\sigma\tau, {\bm k}' \sigma' \tau' \rangle   \nonumber\\
+ \langle{\bm k}\sigma \tau, {\bm k}'\sigma' \tau' |v_{\rm ex}|
{\bm k}'\sigma \tau, {\bm k}\sigma' \tau' \rangle], \label{eq1} 
\end{align}
where $|{\bm k}\sigma \tau\rangle$ are plane waves, $v_{\rm d}$ and $v_{\rm ex}$ 
are the direct and exchange terms of the in-medium NN interaction. 

We have chosen for the present mean-field study several versions of the density 
dependent NN interaction based on the G-matrix M3Y interaction \cite{Anan83}, which 
were used successfully in the HF studies of NM \cite{Loan11,Tan16} and the folding 
model analysis of the nucleus-nucleus scattering \cite{Kho97,Kho00}. Explicitly, the 
density dependent NN interaction is just the original M3Y interaction supplemented 
by a realistic density dependence 
\begin{align}
 v_{\rm d(ex)}(n_{\rm b}, r)=  F_{00}(n_{\rm b})v_{00}^{\rm d(ex)}(r) + 
 F_{10}(n_{\rm b}) v_{\rm 10}^{\rm d(ex)}(r)({\bm\sigma}{\bm\sigma}') \nonumber \\
 +F_{01}(n_{\rm b}) v_{01}^{\rm d(ex)}(r)({\bm\tau}{\bm\tau}')+F_{11}(n_{\rm b}) 
v_{11}^{\rm d(ex)}(r)({\bm\sigma}{\bm\sigma}') ({\bm\tau}{\bm\tau}'). \label{eq2}
\end{align}
We consider in the present work the \emph{spin-saturated} NM, so that the 
$\sigma$-components of plane waves are averaged out in the HF calculation (\ref{eq1}), 
and only the $v_{00}$ and $v_{01}$ terms of the central interaction (\ref{eq2}) are 
necessary for the determination of the energy density of NM. The radial parts of 
the central interaction (\ref{eq2}) are determined from the spin singlet and triplet 
components of the M3Y interaction \cite{Anan83}, and expressed in terms of three Yukawa 
functions \cite{Kho96} as
\begin{equation}
 v^{\rm d(ex)}_{00(01)}(r)=\sum_{\kappa=1}^3 Y^{\rm d(ex)}_{00(01)}(\kappa)
\frac{\exp(-R_\kappa r)}{R_\kappa r}, \label{eq3}
\end{equation} 
with the Yukawa strengths given explicitly in Table~\ref{t1}. 
\begin{table} [bht]
\caption{Yukawa strengths of the M3Y interaction \cite{Anan83}.}
 \vspace{0cm} \label{t1}
\addtolength{\tabcolsep}{0pt}
\begin{tabular}{cccccc}\hline\hline
$\kappa$ & $R_{\kappa}$ & $Y^{\rm d}_{\rm 00}(\kappa)$ & $Y^{\rm d}_{\rm 01}(\kappa)$ 
& $Y^{\rm ex}_{\rm 00}(\kappa)$ & $Y^{\rm ex}_{\rm 01}(\kappa)$ \\
& (fm$^{-1}$) & (MeV) & (MeV) & (MeV) & (MeV)  \\ \hline
1 & 4.0 & 11061.625 &  313.625 & -1524.25 & -4118.0 \\
2 & 2.5 & -2537.5   &  223.5 & -518.75 & 1054.75  \\
3 & 0.7072 & 0.0    &  0.0 &   -7.8474 & 2.6157  \\ \hline\hline
\end{tabular} 
\end{table}
Originally, the \emph{isoscalar} density dependence $F_{00}(n_{\rm b})$ was 
parametrized in Ref.~\cite{Kho97} to correctly reproduce the saturation properties 
of symmetric NM at the saturation baryon density $n_0\approx 0.17$ fm$^{-3}$. About 
20 years after, the \emph{isovector} density dependence $F_{01}(n_{\rm b})$ was 
parametrized to reproduce the microscopic Brueckner-Hartree-Fock results of asymmetric 
NM, with the total strength slightly adjusted by the folding model description 
of the charge exchange $(p,n)$ reaction to the isobar analog states in finite 
nuclei \cite{Kho07,Kho14}. Note that the CDM3Y8 version of the density dependent 
NN interaction has been parametrized recently in the mean-field study of the 
spin-polarized NS matter \cite{Tan20}, including the density dependence of the 
spin- and spin-isospin terms of the interaction (\ref{eq2}). Thus, the following 
density dependent functional is used for the interaction (\ref{eq2})  
\begin{equation}
F_{st}(n_{\rm b}) = C_{st}\big[1+\alpha_{st}\exp(-\beta_{st}n_{\rm b}) 
+ \gamma_{st} n_{\rm b}\big],  \label{eq4}
\end{equation}
and all the parameters are given explicitly in Table~\ref{t2}. 
\begin{table*} [bht]
\caption{Parameters of 6 versions of the density dependent NN interaction 
(\ref{eq2})-(\ref{eq4}) used in the present HF calculation. The parameters 
of $F_{00}(n_{\rm b})$ were adjusted to reproduce the binding energy $E_0\approx 15.8$ 
MeV and different values of the nuclear incompressibility $K$ of symmetric NM at 
the saturation density $n_0\approx 0.17$ fm$^{-3}$. The parameters of $F_{01}(n_{\rm b})$ 
were fine tuned to give $E_{\rm sym}(n_0)\approx 30$ MeV and $L\approx 50$ MeV at $n_0$, 
and a good agreement of the calculated $E_{\rm sym}(n_{\rm b})$ with the ab-initio 
results \cite{APR,MMC} at higher densities.} 
\begin{center}
\vspace{0.cm}\addtolength{\tabcolsep}{0pt}
\begin{tabular}{cccccccccc}\hline\hline 
Interaction & $C_{00}$& $\alpha_{00}$& $\beta_{00}$ & $\gamma_{00}$ & $C_{01}$ & 
 $\alpha_{01}$ & $\beta_{01}$ & $\gamma_{01}$ & $K$  \\
 & & & (fm$^3$) & (fm$^3$) & & & (fm$^3$)& (fm$^3$) & (MeV) \\ \hline
CDM3Y3 & 0.2985 & 3.4528 & 2.6388 & -1.5 & 0.2342 & 5.3336 & 6.4738 & 4.3172 & 217 \\
CDM3Y4 & 0.3052 & 3.2998 & 2.3180 & -2.0 & 0.2129 & 6.3581 & 7.0584 & 5.6091 & 228 \\
CDM3Y5 & 0.2728 & 3.7367 &  1.8294 & -3.0 &  0.2204 & 6.6146 & 7.9910 & 6.0040 & 241 \\
CDM3Y6 & 0.2658 & 3.8033 & 1.4099 & -4.0 & 0.2313 & 6.6865 & 8.6775 & 6.0182 & 252 \\
CDM3Y8 & 0.2658 & 3.8033 & 1.3035 & -4.3 & 0.2643 & 6.3836 & 9.8950& 5.4249 & 257 \\
BDM3Y1 & 1.2521 & 0.0 & 0.0 & -1.7452 & 0.2528 &  7.6996 & 11.0386 & 6.3568 & 270 \\ 
\hline\hline \label{t2}
\end{tabular} 
\end{center}
\end{table*}
The parameters 
of $F_{00}(n_{\rm b})$ of the CDM3Y8 version are slightly fine tuned in the present study
to reproduce the same binding energy $E_0\approx 15.8$ MeV at the saturation density 
$n_0\approx 0.17$ fm$^{-3}$ as those obtained with the other 5 versions of the 
density dependent NN interaction that were originally parametrized in Ref.~\cite{Kho97}. 
The parameters of $F_{01}(n_{\rm b})$ of all 6 versions are also readjusted in the 
present work to reduce the uncertainty at high baryon densities, and to reach a better 
agreement of $E_{\rm sym}(n_{\rm b})$ given by the HF calculation with the ab-initio 
results \cite{APR,MMC} at $n_{\rm b}>n_0$.  
The total HF energy density (\ref{eq1}) of the asymmetric NM is then obtained as
\begin{equation}
\mathcal{E}=\frac{3\hbar^2}{10m}[k^2_{F_{\rm n}}n_{\rm n}+k^2_{F_{\rm p}}n_{\rm p}]
 +F_{00}(n_{\rm b})\mathcal{E}_{00}+F_{01}(n_{\rm b})\mathcal{E}_{01}, \label{eq5} 
\end{equation} 
where $m$ is the nucleon mass, and the potential energy density of NM is determined
with
\begin{align}
& \mathcal{E}_{00}=\frac{1}{2}\left[n_{\rm b}^2 J^{\rm d}_{00}+ 
 \int A^2_{00} v^{\rm ex}_{00}(r) d^3r\right], \nonumber\\
& \mathcal{E}_{01}=\frac{1}{2}\left[n_{\rm b}^2 J^{\rm d}_{01}\delta^2 + 
 \int A^2_{01} v^{\rm ex}_{01}(r) d^3r\right]. \label{eq6} 
\end{align}
Here $J^{\rm d}_{00(01)}=\displaystyle\int v^{\rm d}_{00(01)}(r)d^3r$ is the volume 
integral of the direct interaction, and the exchange integrals are evaluated with
\begin{equation}
 A_{00(01)}=n_{\rm n}\widehat{j}_1(k_{F_{\rm n}}r)\pm 
 n_{\rm p} \widehat{j}_1(k_{F_{\rm p}}r), \label{eq7}
\end{equation}
where $\widehat{j}_1(x)=3j_1(x)/x$, and $j_1(x)$ is the first-order spherical Bessel 
function. The nucleon Fermi momentum is determined as 
$k_{F_{\rm n(p)}}=[3\pi^2n_{\rm n(p)}]^{1/3}$. One can see that the nonzero neutron-proton
asymmetry ($\delta\neq 0$ or $n_{\rm n}\neq n_{\rm p}$) results on a nonzero contribution 
from $\mathcal{E}_{01}$ term to the total NM energy density (\ref{eq5}), which is the
case for both the crust and core of NS. 

Dividing the total NM energy density (\ref{eq5}) by the baryon number density, we 
obtain the NM energy per baryon $E/A$ which can be expanded over the neutron-proton 
asymmetry $\delta$ as
\begin{equation}
\frac{\mathcal{E}}{n_{\rm b}}\equiv\frac{E}{A}(n_{\rm b},\delta)=
\frac{E}{A}(n_{\rm b},\delta=0)+E_{\rm sym}(n_{\rm b})\delta^2+O(\delta^4)+..., 
 \label{eq8}
\end{equation}
where $E_{\rm sym}(n_{\rm b})$ is the {\bf nuclear symmetry energy}. 
\begin{figure}[bht!]\vspace{-1.5cm}
\includegraphics[angle=0,width=0.53\textwidth]{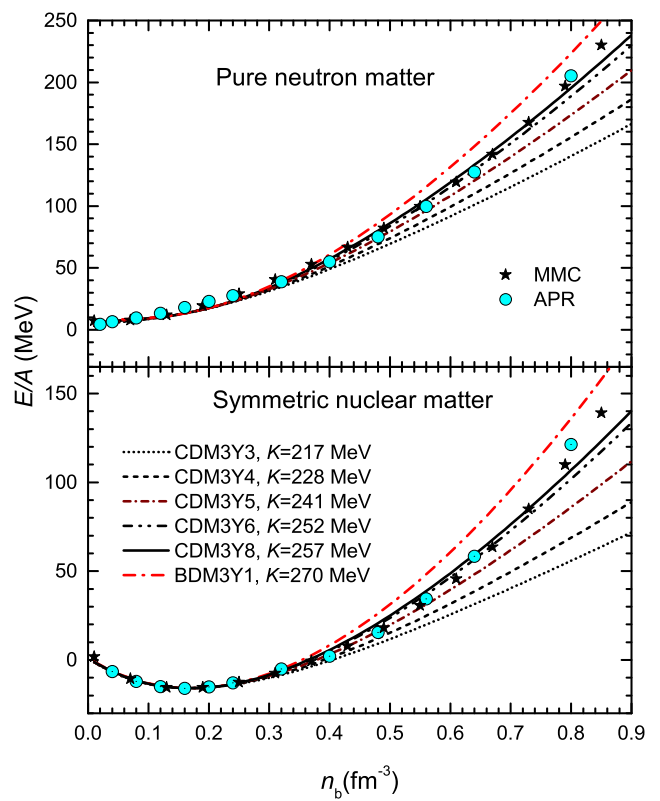}\vspace{-0.7cm}
 \caption{The energy per baryon of symmetric NM (lower panel) and pure neutron matter 
(upper panel) given by the HF calculation, using 6 versions of the density dependent 
NN interaction (\ref{eq2})-(\ref{eq4}). The circles and stars are the results 
of the ab-initio calculation by Akmal, Pandharipande and Ravenhall (APR) \cite{APR} 
and microscopic Monte Carlo (MMC) calculation by Gandolfi {\it et al.} \cite{MMC}, 
respectively. $K$ is the nuclear incompressibility (\ref{eq10}) of symmetric NM at the 
saturation density $n_0\approx 0.17$ fm$^{-3}$.} \label{f1}
\end{figure}
The contribution from $O(\delta^4)$ and higher-order terms in Eq.~(\ref{eq8}) is 
proven to be small and can be neglected in the \emph{parabolic} approximation \cite{Kho96}, 
where $E_{\rm sym}(n_{\rm b})$ equals the energy required per baryon to change 
the symmetric NM (with $\delta=0$) to the pure neutron matter (with $\delta=1$). 
Thus, the two key quantities characterizing EOS of NM are the energy of symmetric 
NM and nuclear symmetry energy. At the saturation density $n_0\approx 0.17$ fm$^{-3}$, 
these two quantities are constrained by the empirical values inferred from the structure 
studies of finite nuclei 
\begin{equation}
 E_0=\frac{E}{A}(n_0,\delta=0)\approx -15.8\ {\rm MeV},\  
 E_{\rm sym}(n_0)\approx 30\ {\rm MeV}. \nonumber
\end{equation} 
However, their behavior at higher baryon densities of NM ($n_{\rm b}>n_0$) remains  
less certain because of the difficulties for the terrestrial nuclear physics
experiments to access the high-density NM. The pressure of baryon matter inside 
the high-density medium of NS is determined as
\begin{align}
 P(n_{\rm b},\delta)=n_{\rm b}^2 \frac{\partial}{\partial n_{\rm b}}
 \left[\frac{E}{A}(n_{\rm b},\delta)\right]=P(n_{\rm b},\delta=0) \nonumber\\
 +P_{\rm sym}(n_{\rm b})\delta^2+O(\delta^4)+..., \label{eq9}
\end{align}
which is expanded over $\delta$ in the same parabolic approximation \cite{Kho96}  
as in Eq.~(\ref{eq8}). The total energy of symmetric NM and nuclear symmetry 
energy at high baryon densities correlate strongly with the nuclear incompressibility 
$K$ and slope parameter of the symmetry energy $L$, respectively, determined 
at the saturation density as 
\begin{equation}
 K=9 \frac{\partial P(n_{\rm b},\delta=0)}{\partial n_{\rm b}}
 \Bigr\vert_{n_{\rm b}\to n_0},\ L=3n_{\rm b}\frac{\partial E_{\rm sym}(n_{\rm b})}
 {\partial n_{\rm b}}\Bigr\vert_{n_{\rm b}\to n_0}.  \label{eq10}
\end{equation}
The $K$ and $L$ values are strongly sensitive to the EOS of NM. The compressibility
$K$ has been a key research topic of numerous structure studies of nuclear monopole 
excitation (see, e.g., Ref.~\cite{Garg18} and references therein) as well as 
studies of the heavy-ion (HI) collisions \cite{Schu96} and nucleus-nucleus 
scattering \cite{Kho07r}. These researches have pinned down this quantity to 
$K\approx 240\pm 20$ MeV. The present HF calculation (\ref{eq10}) of NM using 6 
different versions of the density dependent NN interactions (\ref{eq2})-(\ref{eq4}) 
tabulated in Table~\ref{t2} gives $K\approx 217 - 270$ MeV, which agree well 
with the empirical value.

The HF results for the energy per baryon of symmetric NM and pure neutron 
matter and are plotted in Fig.~\ref{f1}, in comparison with the \emph{ab-initio} 
results of the many-body variational calculation by Akmal {\it et al.} \cite{APR} 
using the Argonne-AV18 NN potential, and microscopic Monte Carlo calculation 
of NM by Gandolfi {\it et al.} \cite{MMC} using the Argonne-AV6' NN potential 
supplemented by an empirical density dependence. The sensitivity of the EOS to the 
$K$ value is well illustrated in Fig.~\ref{f1}, and the HF results given by the 
CDM3Y6 and CDM3Y8 interactions (associated with $K\approx 252$ and 257 MeV, 
respectively) agree nicely with the ab-initio results \cite{APR,MMC} at high 
baryon densities. The effect caused by the difference in the $K$ values shows 
up also in the macroscopic properties of NS (see the discussion below 
in Sec.~\ref{sec4}).   

\begin{figure}[bht!]\vspace{-1.0cm}\hspace{-0.6cm}
\includegraphics[angle=0,width=0.57\textwidth]{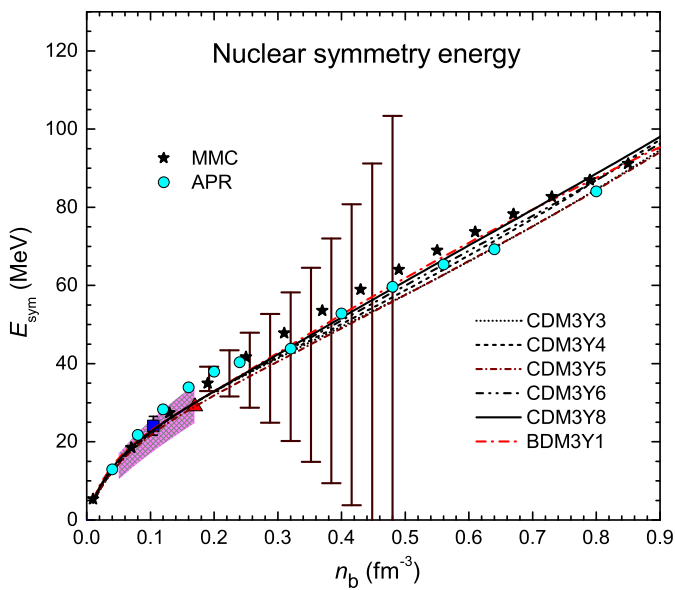}\vspace{0cm}
 \caption{The nuclear symmetry energy $E_{\rm sym}(n_{\rm b})$ given by the HF 
calculation using 6 versions of the density dependent NN interaction (\ref{eq2})-(\ref{eq4}). 
The shaded region is the range constrained by the HI data \cite{Tsang11,Ono03}, the square 
and triangle are values suggested by the structure studies \cite{Tri08,Fur02,Dong18}. 
The vertical bars are the empirical range obtained at the 90\% confidence level 
from the Bayesian analysis \cite{Xie19} of the NS radius $R_{1/4}$ versus 
the GW170817 constraint \cite{Abbott18}. The same denotations as in Fig.~\ref{f1} 
for the circles and stars.} \label{f2} 
\end{figure}  
At variance with the nuclear incompressibility $K$ that helps to constrain 
the behavior of the EOS of symmetric NM at high densities, it is more 
difficult to constrain the density dependence of the nuclear symmetry energy 
$E_{\rm sym}$ at densities $n_{\rm b}>n_0$. In fact, $E_{\rm sym}$ has been well 
constrained at low baryon densities by the (isospin dependent) data of the HI 
fragmentation \cite{Tsang11,Ono03}, and by the structure studies of the giant 
dipole resonances \cite{Tri08}, neutron skins \cite{Fur02,Dong18}, and nuclear mass 
differences \cite{Fan14}. The HF results for $E_{\rm sym}$  at low baryon densities 
agree well with the empirical data (see Fig.~\ref{f2}). The behavior of the nuclear 
symmetry energy at higher baryon densities ($n_{\rm b}>n_0$) remains poorly known. 
One can see in Fig.~\ref{f2} that the HF results obtained for $E_{\rm sym}$ at 
$n_{\rm b}>n_0$ with the revised isovector density dependence of the interaction  
agree well with the ab-initio results \cite{APR,MMC}, and the slope parameter 
of the nuclear symmetry energy $L\approx 50$ MeV is obtained with all 6 versions 
of the density dependent NN interaction (see Table~\ref{t2}). Recently, the symmetry 
energy at baryon densities up to $3n_0$ was inferred from the Bayesian analysis 
of the correlation of different EOS's of np$e\mu$ matter and the predicted radius 
$R_{1/4}$ of NS with mass $M=1.4~M_{\odot}$ versus the GW170817 constraint imposed 
by the tidal deformability of NS. The empirical $E_{\rm sym}$ values suggested 
for baryon densities approaching $3 n_0$ at the 90\% confidence level \cite{Xie19} 
are shown as the vertical bars in Fig.~\ref{f2}, and the uncertainty still remains 
so large that the results of the HF and ab-initio calculations are all enclosed 
inside the empirical boundaries, over the whole range of baryon density. 

\section{$\bm{\beta}$-stable np$\bm{e\mu}$ matter of neutron star}
\label{sec3}
The HF model (\ref{eq1}) describes the ideal NM that contains nucleons only. In fact,
the NS matter contains significant lepton fraction in both the crust and uniform core,
and a realistic EOS of NS matter should include the lepton contribution. In the present
work, we use the EOS of the NS crust obtained by Douchin {\it et al.} \cite{Dou00,Dou01} 
in the Compress Liquid Drop Model based on the SLy4 interaction between nucleons 
\cite{Cha98}, with fully ionized atoms and a degenerate Fermi gas of free electrons. 
At the edge density $n_{\rm edge}\approx 0.076$ fm$^{-3}$, a weak first-order 
phase transition takes place between the NS crust and uniform core of NS \cite{Dou00}. 
At baryon densities $n_{\rm b}\gtrsim n_{\rm edge}$ the core of NS is described as a 
homogeneous matter of neutrons, protons, electrons and negative muons ($\mu^-$ appear 
at $n_{\rm b}$ above the muon threshold density $\mu_e>m_\mu c^2\approx 105.6$ MeV). 
The total mass-energy density $\mathcal{E}$ of the np$e\mu$ matter inside NS core is 
determined as
\begin{eqnarray}
\mathcal{E}(n_{\rm n},n_{\rm p},n_e,n_\mu)&=&  
\mathcal{E}_{\rm HF}(n_{\rm n},n_{\rm p})+n_{\rm n}m_{\rm n}c^2
+n_{\rm p}m_{\rm p}c^2 \nonumber\\
&+&\mathcal{E}_e(n_e)+\mathcal{E}_\mu(n_\mu) , \label{eq11} 
\end{eqnarray}
where $\mathcal{E}_{\rm HF}(n_{\rm n},n_{\rm p})$ is the HF energy density (\ref{eq3}) 
of asymmetric NM, $\mathcal{E}_e$ and $\mathcal{E}_\mu$ are the energy densities 
of electrons and muons given by the relativistic Fermi gas model \cite{Shapiro}. In 
such a Fermi gas model, the lepton number densities $n_e$ and $n_\mu$ are determined 
from the charge neutrality condition ($n_{\rm p}=n_e+n_\mu$), and the $\beta$-equilibrium 
of (neutrino-free) NS matter is sustained by the balance of the chemical potentials 
\begin{equation}
\mu_{\rm n} = \mu_{\rm p}+\mu_e\ \ \text{and}\ \ \mu_e=\mu_\mu,\ 
\text{where}\ \mu_j=\frac{\partial\mathcal{E}_j}{\partial n_j}. \label{eq12}
\end{equation} 
The fractions of the constituent particles in NS matter are determined at the given 
baryon density $n_{\rm b}$ as $x_j=n_j/n_{\rm b}$. Below the muon threshold density 
($\mu_e<m_\mu c^2\approx 105.6$ MeV) the charge neutrality condition leads to the 
following relation \cite{Bom01}
\begin{equation}
 3\pi^2(\hbar c)^3n_{\rm b}x_{\rm p}-\hat\mu^3=0,\ {\rm where}\
 \hat\mu=\mu_{\rm n}-\mu_{\rm p}=2\frac{\partial\mathcal{E}_{\rm HF}}
 {\partial\delta}. \label{eq13}
\end{equation}
The density dependence of the proton fraction in the $\beta$-equilibrium,
$x_{\rm p}(n_{\rm b})$, can then be obtained from the solution of Eq.~(\ref{eq13}).
If we assume the parabolic approximation and neglect the contribution from
higher-order terms in (\ref{eq8}), then $x_{\rm p}(n_{\rm b})$ is given by the
solution of the well-known equation \cite{Bom01}
\begin{equation}
 3\pi^2(\hbar c)^3n_{\rm b}x_{\rm p}-[4E_{\rm sym}(n_{\rm b})(1-2x_{\rm p})]^3=0, 
 \label{eq14}
\end{equation}
which shows the important role of the nuclear symmetry energy in the determination 
of the proton abundance in NS matter.
\begin{figure}[bht]\vspace{-1cm}
\hspace{-1cm}\includegraphics[width=0.63\textwidth]{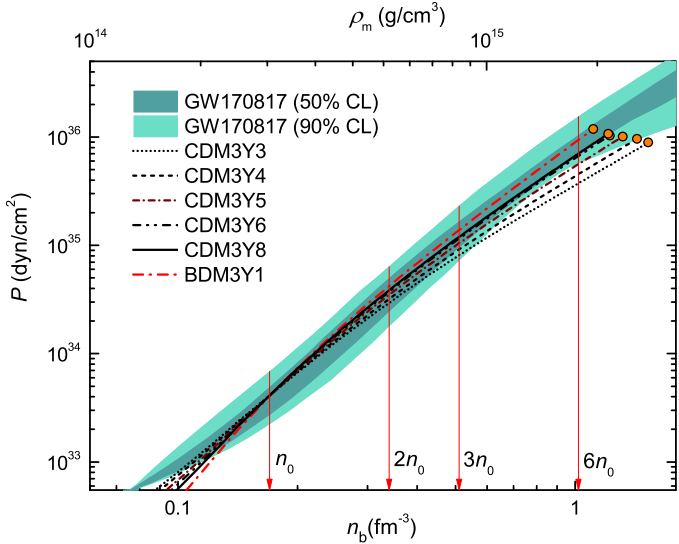}\vspace*{0cm}
 \caption{The total pressure of NS matter (\ref{eq16}) over the whole range of the 
rest-mass ($\rho_{\rm m}$) and baryon-number ($n_{\rm b}$) densities, given by the HF 
calculation (\ref{eq11})-(\ref{eq16}) using 6 versions of the density dependent NN
interaction (\ref{eq2})-(\ref{eq4}). The dark and light shaded regions enclose 
the empirical pressure given by the ``spectral" EOS inferred from the Bayesian analysis 
of the GW170817 data with the 50\% and 90\% confidence levels, respectively, keeping 
the lower limit of the maximum NS mass at $1.97~M_\odot$ \cite{Abbott18}. The circles 
are the pressure $P(n_{\rm c})$ at the maximum central density $n_{\rm c}$ obtained 
with each interaction (see Table~\ref{t3}).} \label{f3}
\end{figure}

As baryon densities exceeding the muon threshold density, where $\mu_e >
m_\mu c^2\approx 105.6$ MeV, it is energetically favorable for electrons
to convert to negative muons, and the charge neutrality condition leads to the
relation \cite{Bom01}
\begin{equation}
 3\pi^2(\hbar c)^3n_{\rm b}x_{\rm p}-\hat\mu^3-
 [\hat\mu^2-(m_\mu c^2)^2]^{3/2}\theta(\hat\mu-m_\mu c^2)=0, \label{eq15}
 \end{equation}
where $\theta(x)$ is the Heaviside step function. Based on the solutions of
Eqs.~(\ref{eq11})-(\ref{eq15}), the EOS of the $\beta$-stable np$e\mu$ matter  
is fully determined by the mass-energy density $\rho(n_{\rm b})$ and total 
pressure $P(n_{\rm b})$ of NS matter  
\begin{equation}
\rho(n_{\rm b})=\mathcal{E}(n_{\rm b})/c^2,\ P(n_{\rm b})=n_{\rm b}^2
{\frac{\partial}{{\partial n_{\rm b}}}}\left[\frac{\mathcal{E}_{\rm HF}(n_{\rm b})}
 {n_{\rm b}}\right]+P_e+P_\mu. \label{eq16}
\end{equation}
We show in Fig.~\ref{f3} the total pressure of NS matter $P(n_{\rm b})$ given by the 
HF calculation using 6 versions of the density dependent NN interaction, over baryon 
densities up to above $6n_0$, in comparison with the empirical pressure given by the 
``spectral" EOS inferred from the Bayesian analysis of the GW170817 data at the 
50\% and 90\% confidence levels \cite{Abbott18}. 
In general, our mean-field results agree reasonably with the empirical pressure over 
baryon densities up to 3$n_0$. The empirical pressure keeps steadily rising up at 
the highest densities in the center of NS, and this trend is due to the fact that 
sampling of the spectral EOS has been constrained \cite{Abbott18} by the maximum NS 
mass $M_{\rm max}\gtrsim 1.97~M_\odot$, based on the observation of the second 
PSR J$0348+0432$ with $M\approx 2.01\pm 0.04~M_\odot$ \cite{Sci340}. Excepting the
results obtained with the CDM3Y3 and CDM3Y4 interactions, the pressure of NS matter 
given by the other 4 versions of the density dependent NN interaction at high baryon 
densities follows closely the empirical trend up to the corresponding maximum 
central densities $n_{\rm c}$ (see Fig.~\ref{f3} and Table~\ref{t3}). It is obvious 
that the lower pressure at high densities given by the CDM3Y3 and CDM3Y4 interactions 
is due to the smaller $K$ values associated with these interactions. With the larger 
$K$ values, the NS matter generated by the CDM3Y5, CDM3Y6, CDM3Y8 and BDM3Y1 interactions 
can be compressed to have higher pressure at high baryon densities, in agreement 
with the empirical pressure inferred from the GW170817 data. Because all 6 versions 
of the density dependent NN interaction were adjusted to obtain the same saturation
energy $E_0\approx 15.8$ MeV and $L\approx 50$ MeV at the saturation density $n_0$, 
the NM pressure (9) is determined mainly by its symmetry term which is directly 
proportional to $L$, and the $P(n_0)$ values shown in Fig.~\ref{f3} are nearly 
the same. At low baryon densities $n_{\rm b}<n_0$, the slope of the energy of symmetric
NM is opposite compared to that at $n_{\rm b}>n_0$ (see Fig.~\ref{f1}), leading 
to the decrease of $P(n_{\rm b})$ with the increasing incompressibility $K$ 
at subsaturation densities shown in Fig.~\ref{f3}. 

\begin{figure}[bht]\vspace{-0.5cm}
\hspace{-1cm}\includegraphics[width=0.62\textwidth]{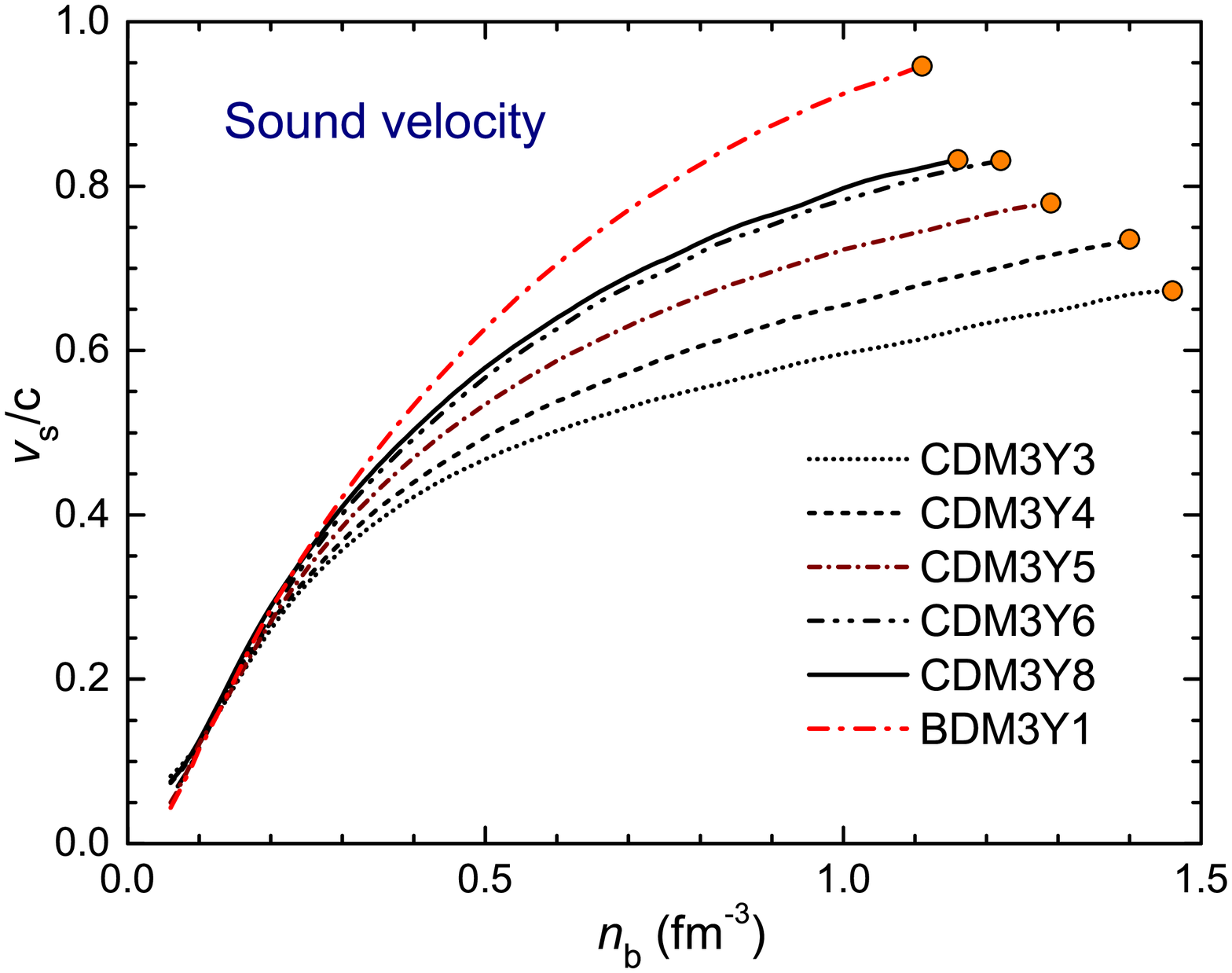}\vspace*{-0.5cm}
 \caption{The adiabatic sound velocity (\ref{eq16s}) in the NS medium given by 
6 versions of the density dependent NN interaction (\ref{eq2})-(\ref{eq4}). 
The circles are the $v_{\rm s}$ values obtained at the maximum central densities 
$n_{\rm c}$ with each interaction.} 
\label{f3s}
\end{figure}

Because the total pressure of NS matter is directly proportional to the total 
mass density, it is of interest to consider the \emph{causality} condition \cite{Bom01} 
that implies the adiabatic sound velocity in the stellar medium to be \emph{subluminal}
\begin{equation}
 v_{\rm s}=\sqrt{\frac{dP(\rho)}{d\rho}}< {\rm c},  \label {eq16s}
\end{equation}
where $P(\rho)$ is the total pressure of NS matter deduced from the relation 
(\ref{eq16}) as function of the total mass density $\rho$. The sound velocity 
$v_{\rm s}$ given by the total pressure obtained with 6 versions of the density 
dependent NN interaction at different baryon number densities is shown in 
Fig.~\ref{f3s}, and one can see that the causality condition is well satisfied 
by the present EOS of NS matter. With about the same $K$ values associated with 
the CDM3Y6 and CDM3Y8 interactions, these two versions of the interaction give 
more or less the same $v_{\rm s}$ values over the whole range of baryon density. 
The higher central pressure $P(\rho)$ associated with the higher nuclear incompressibility 
$K$ corresponds also to the higher sound velocity. Given the recent discussion on the  
${\rm c}/\sqrt{3}$ bound of the sound velocity \cite{Bed15}, it is complimentary 
to note that the EOS's obtained with the CDM3Y6, CDM3Y8, and BDM3Y1 interactions gives 
$v_{\rm s}\gtrsim {\rm c}/\sqrt{3}$ at baryon densities $n_{\rm b}\gtrsim 3n_0$ 
where the corresponding pressures of NS matter also agree nicely with 
the empirical data \cite{Abbott18} (see Fig.~\ref{f3}).    

\section{Tidal deformability, mass and radius of neutron star}
\label{sec4}
An exciting effect observed by the GW170817 merger of two neutron stars is the
tidal deformation of NS induced by the strong gravitational field, which enhances the
GW emission and accelerates the decay of the quasicircular inspiral \cite{Abbott18}.   
We give here a brief account of the tidal deformability of a static spherical star  
being in the hydrostatic equilibrium \cite{Hinderer08,Hinderer10}. If exposed to the 
gravitational field created by the attraction of the companion star in a binary system, 
this star is tidally deformed and gains a nonzero quadrupole moment $Q_{ij}$ that is 
directly proportional to the field strength $E_{ij}$ 
\begin{equation}
 Q_{ij} = -\lambda E_{ij}. \label{eq17}
\end{equation}
$\lambda$ characterizes the star response to the gravitational field and is dubbed 
as the \emph{tidal deformability} or quadrupole polarizability of star. In the General
Relativity, $\lambda$ is related to the $l=2$ tidal Love number $k_2$ 
\cite{Hinderer08} as
\begin{equation}
 \lambda = \frac{2}{3G}k_2 R^5, \label{eq18}
\end{equation}
where $R$ and $G$ are the star radius and gravitational constant, respectively. 
For the discussion on the sensitivity to the EOS of NS matter, it is convenient 
to consider the dimensionless tidal deformability parameter $\Lambda$ \cite{Abbott18}
expressed in terms of the compactness $C$ of star with mass $M$ and radius $R$ as 
\begin{equation}
 \Lambda=\frac{2}{3}k_2C^{-5}, \ {\rm with}\ C=\frac{GM}{Rc^2}. \label{eq19}
\end{equation}
Using the linearized Einstein equation, the Love 
number $k_2$ can be expressed \cite{Hinderer08,Hinderer10} as 
\begin{align}
k_2=\frac{8C^5}{5}(1-2C)^2 [2+2C(y-1)-y] \nonumber \\ 
\times \{ 2C [6-3y+3C(5y-8)]\nonumber \\ 
 +4C^3[13-11y+C(3y-2)+2C^2(1+y)] \nonumber \\
 + 3(1-2C)^2 [2-y+2C(y-1)] \ln(1-2C) \}, \label{eq20}
\end{align}
where $y=RH'(R)/H(R)$, and function $H(r)$ is related to the nonzero component 
of the metric perturbation of the stress-energy tensor. $H(r)$ and its radial 
derivative $H'(r)$ can be determined \cite{Hinderer08,Hinderer10} from the 
solution of the following differential equation
\begin{align}
 H''(r)+H'(r)\left[1-\frac{2G\mathcal{M}(r)}{c^2r}\right]^{-1}
 \Bigg\{\frac{2}{r} - \frac{2G\mathcal{M}(r)}{c^2r^2} \nonumber \\
 - \frac{4\pi G}{c^4} r\left[\rho(r)-P(r)\right] \Bigg\} 
 + H(r)\left[1-\frac{2G\mathcal{M}(r)}{c^2r}\right]^{-1} \nonumber \\
 \Bigg\{\frac{4\pi G}{c^4}\left[5\rho(r)+9P(r)+\frac{d\rho(r)}{dP(r)}
 \left(\rho(r)+ P(r))\right)\right]-\frac{6}{r^2} \nonumber \\
 -4\left[1-\frac{2G\mathcal{M}(r)}{c^2r}\right]^{-1}
 \left[\frac{G\mathcal{M}(r)}{c^2r^2}+
\frac{4\pi G}{c^4}rP(r)\right]^2\Bigg\}=0. \label{eq21}
\end{align} 
Here $r$ is the radial coordinate in Schwarzschild metric, and $\mathcal{M}(r)$ 
is the gravitational mass enclosed within the sphere of radius $r$. 
$\rho(r)$ and $P(r)$ are the mass-energy density and pressure of star matter 
at radius $r$, respectively. In the present work, Eq.~(\ref{eq21}) is solved 
using the fourth-order Runge-Kutta method and integrated together with the  
Tolman-Oppenheimer-Volkoff (TOV) equations 
\begin{eqnarray}
\frac{d P(r)}{dr}&=& -\frac{G \rho(r)\mathcal{M}(r)}{c^2 r^2} 
 \left[ 1+\frac{P(r)}{\rho(r)}\right] 
\left[1+\frac{4 \pi P(r) r^3}{c^2 \mathcal{M}(r)} \right] \nonumber\\
&&\ \quad \times \left[ 1-\frac{2G \mathcal{M}(r)}{c^2 r} \right]^{-1} , \nonumber \\
d\mathcal{M}(r) &=& 4\pi r^2 \rho(r)dr. \label{eq22} 
\end{eqnarray}
Eqs.~(\ref{eq21}) and (\ref{eq22}) are integrated from the star center to
its surface, with the star radius $R$ given by $P(R)=0$. 
Other boundary conditions are 
\begin{align}
\mathcal{M}(0)=0, \ \rho(0)=\rho_{\rm c}, \ P(0)=P_{\rm c}, \nonumber \\  
H(0)=0,\ H'(0)=0, \ \mathcal{M}(R)=M. \label{eq23}  
\end{align}

\begin{figure}[bht]\vspace{-0.5cm}
\hspace{-0.5cm}\includegraphics[width=0.6\textwidth]{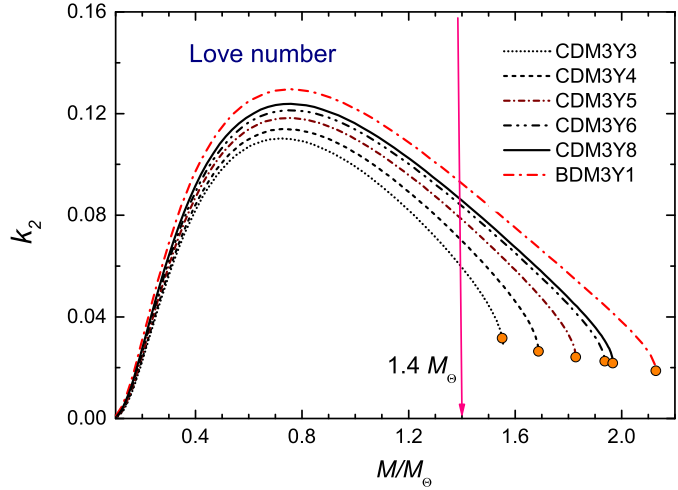}\vspace*{-0.5cm}
\caption{The Love number (\ref{eq20}) at different masses of NS given by the EOS 
of $\beta$-stable np$e\mu$ matter obtained with 6 versions of the density dependent 
NN interaction (\ref{eq2})-(\ref{eq4}). The arrow crosses the $k_2$ values at 
$M=1.4~M_\odot$, and the circles are those obtained at the corresponding maximum 
central densities $n_{\rm c}$.} \label{f4}
\end{figure}
\begin{figure}[bht]\vspace{-0.5cm}
\hspace{-1.0cm}\includegraphics[width=0.63\textwidth]{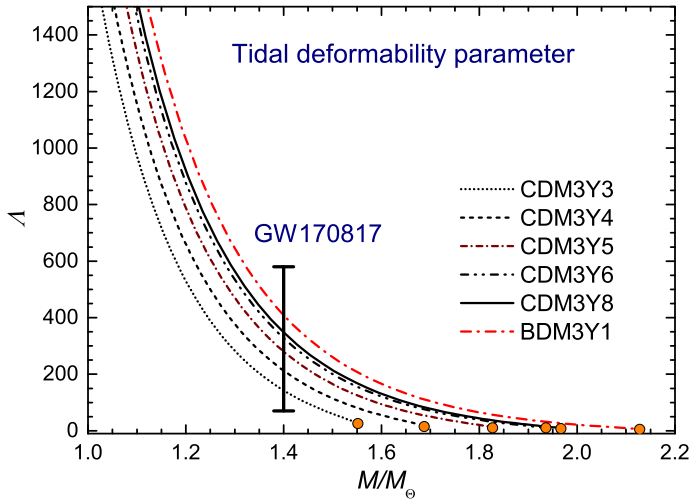}\vspace*{-0.5cm}
 \caption{The tidal deformability parameter (\ref{eq19}) given by the EOS of 
$\beta$-stable np$e\mu$ matter obtained with 6 versions of the density dependent 
NN interaction (\ref{eq2})-(\ref{eq4}). The vertical bar is the empirical tidal 
deformability at $M=1.4~M_\odot$ inferred from the Bayesian analysis of the GW170817 
data at the 90\% confidence level \cite{Abbott18}, and the circles are $\Lambda$ values 
obtained at the corresponding maximum central densities $n_{\rm c}$.} \label{f5}
\end{figure}
\begin{table*} [bht]
\begin{center}
\caption{Properties of NS given by the EOS of $\beta$-stable np$e\mu$ matter obtained 
with 6 versions of the density dependent NN interaction (\ref{eq2})-(\ref{eq4}). $M$ 
and $R$ are the maximum gravitational mass and radius of NS, respectively; $n_{\rm c}$ 
and $P_{\rm c}$ are the baryon number density and total pressure in the center of NS. 
$P(2n_0)$ and $P(6n_0)$ are the total pressure at twice and six times the saturation density, 
respectively. $R_{1.4},\ {k_2}_{(1.4)},\ \Lambda_{1.4}$, and $C_{1.4}$ are the radius, 
Love number, tidal deformability parameter, and compactness of NS with $M=1.4~M_\odot$, respectively.} \vspace{0cm} 
\addtolength{\tabcolsep}{0pt}
\begin{tabular}{ccccccccccc}\hline\hline\label{t3}
 EOS & $M$ & $R$ & $n_c$ & $P_c/10^{35}$ & $P(2n_0)/10^{34}$  & $P(6n_0)/10^{35}$ 
 & $R_{1.4}$ & ${k_2}_{(1.4)}$ & $\Lambda_{1.4}$ & $C_{1.4}$ \\
 & ($M_\odot$) & (km) & (fm$^{-3}$) & (dyn/cm$^2$) & (dyn/cm$^2$) & (dyn/cm$^2$) 
 & (km) & & & \\ \hline
 CDM3Y3 & 1.55 & 9.3 & 1.53 & 8.9 & 2.9 & 3.7 & 10.6 & 0.059 & 142 & 0.194 \\
 CDM3Y4 & 1.69 & 9.6 & 1.43 & 9.6 & 3.2 & 4.6 & 11.2 & 0.070 & 212 & 0.186 \\
 CDM3Y5 & 1.83 & 9.9 & 1.32 & 10.1 & 3.5 & 5.6 & 11.5 & 0.078 & 280 & 0.180 \\
 CDM3Y6 & 1.94 & 10.1 & 1.22 & 10.3 & 3.8 & 6.8 & 11.7 & 0.084 & 327 & 0.176\\
 CDM3Y8 & 1.97 & 10.1 & 1.21 & 10.7 & 3.8 & 7.2 & 11.8 & 0.086 & 348 & 0.175 \\ 
 BDM3Y1 & 2.12 & 10.4 & 1.11 & 11.8 & 4.2 & 9.6 & 12.0 & 0.092 & 407 & 0.172 \\
\hline \hline
\end{tabular} \vspace{0.cm}        
\end{center}
\end{table*}
Our mean-field results (\ref{eq19})-(\ref{eq21}) for the Love number and tidal 
deformability of NS are shown in Figs.~\ref{f4} and \ref{f5}, respectively.
The $k_2$ number is reaching its maximum at $M\approx 0.7\sim 0.8~M_\odot$, as
found earlier by Hinderer \emph{et al.} \cite{Hinderer10} using different EOS's 
of NS matter. With the increasing NS mass we observe a sizable difference between 
the $k_2$ values given by 6 versions of the density dependent NN interactions 
(\ref{eq2})-(\ref{eq4}). Such a subtle effect is mainly due to the difference in 
the nuclear incompressibility $K$ given by these interactions (see Table~\ref{t2}).
For NS with mass $M=1.4~M_\odot$, the larger the $R_{1.4}$ radius the larger the 
corresponding $k_2$ number. The tidal deformability of NS with mass up to above 
$2~M_\odot$ is shown in Fig.~\ref{f5}, and $\Lambda$ gradually decreases with 
the increasing NS mass, in a manner similar to that of the Love number shown in 
Fig.~\ref{f4}. The heavier the NS the smaller its tidal deformability, and the 
$\Lambda$ values given by 6 versions of the density dependent NN interaction 
for NS with $M=1.4~M_\odot$ are all inside the empirical range implied by the 
GW170817 data at the 90\% confidence level, $\Lambda_{1.4}\approx 190^{+390}_{-120}$ 
\cite{Abbott18}. The impact by the nuclear incompressibility is significant: 
with $K$ ranging from 217 to 270 MeV the corresponding $\Lambda_{1.4}$ value 
increases nearly by a factor of three (see Table~\ref{t3}).  

\begin{figure}[bht]\vspace{-1.3cm}
\hspace{-0.7cm}\includegraphics[width=0.62\textwidth]{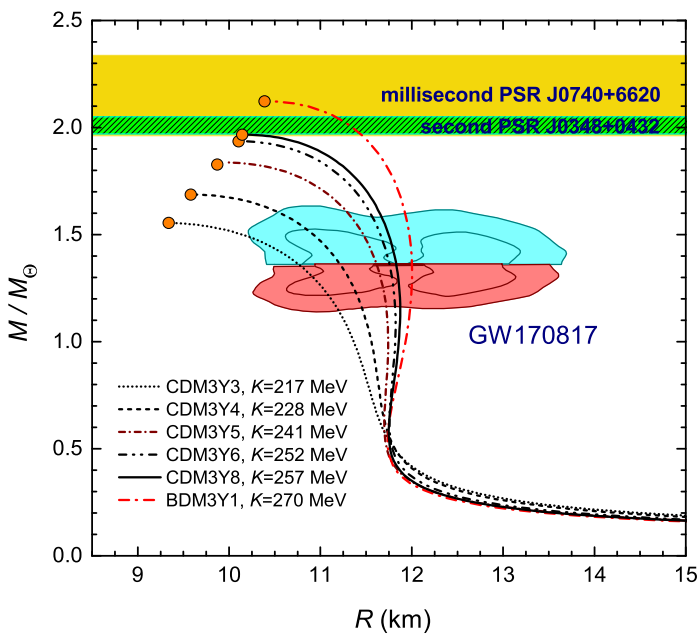}\vspace*{-0.5cm}
\caption{The gravitational mass of NS versus its radius given by the EOS of $\beta$-stable 
np$e\mu$ matter obtained with 6 versions of the density dependent NN interaction 
(\ref{eq2})-(\ref{eq4}). The GW170817 constraint for NS with mass $M=1.4~M_\odot$ 
\cite{Abbott18} is enclosed in the colored contours, and the circles are the $M$-$R$ 
values calculated at the corresponding maximum central densities $n_{\rm c}$. 
The shaded areas enclose the masses of the second PSR J$0348+0432$ \cite{Sci340} 
and millisecond PSR J$0740+6620$ \cite{Cro20}, the heaviest neutron stars observed 
so far.} \label{f6} 
\end{figure}

From the macroscopic properties of NS given by the EOS of $\beta$-stable np$e\mu$ 
matter obtained with 6 versions of the interaction presented in Table~\ref{t3}
one can see a strong impact by the nuclear incompressibility $K$ on the maximum 
gravitational mass $M$ and radius $R$ of NS. Namely, only the $M$ values predicted 
by the CDM3Y6, CDM3Y8, and BDM3Y1 interactions (which are associated with $K=252$, 257,
and 270 MeV, respectively) are close to the lower mass limit of the second 
PSR J$0348+0432$ ($M\approx 2.01\pm 0.04~M_\odot$) \cite{Sci340} and the millisecond 
PSR J$0740+6620$ ($M\approx 2.14^{+0.20}_{-0.18}~M_\odot$) \cite{Cro20}. 
The total pressure $P$ of NS matter given by the CDM3Y6, CDM3Y8, and BDM3Y1 
interactions at twice and six times the saturation density also agree well with 
the empirical pressure deduced from the GW170817 data at the 90\% confidence level, 
$P(2n_0)\approx 3.5^{+2.7}_{-1.7}\times 10^{34}$ and 
$P(6n_0)\approx 9.0^{+7.9}_{-2.6}\times 10^{35}$ dyn/cm$^2$ \cite{Abbott18}. 
The pressure given by the CDM3Y3 and CDM3Y4 interactions (associated with $K\approx 217$ 
and 228 MeV) becomes lower than the empirical pressure at high baryon densities 
$n_{\rm b}>3n_0$ as shown in Fig.~\ref{f3}. The difference in the $K$ values also 
shows up in the calculated NS radius, and $R_{1.4}\approx 11.8$ and 12 km given by 
the EOS obtained with the CDM3Y8 and BDM3Y1 interactions, respectively, is about 
1 km larger than those obtained with the CDM3Y3 and CDM3Y4 interactions (see 
Table~\ref{t3} and Fig.~\ref{f6}). Excepting the results obtained with the CDM3Y3 
interaction, the NS radii predicted by the EOS's obtained with other 5 versions of the 
density dependent NN interaction comply well with the empirical value of $R_{1.4}\approx 
11.75^{+0.86}_{-0.81}$ km deduced recently from a joint analysis \cite{Dietrich20} 
of the two gravitational wave events GW170817 and GW190425, which originated 
from two different NS mergers.  

Thus, the new constraints on the EOS of NS matter at high baryon densities 
\cite{Abbott18,Dietrich20} implied by the extra-terrestrial observations of the NS 
merger and the heaviest pulsars PSR J$0348+0432$ and PSR J$0740+6620$ can be 
used to probe the sensitivity of the EOS of NS matter to the nuclear incompressibility 
$K$. For the illustration, the NS mass versus its radius given by the EOS's of 
$\beta$-stable np$e\mu$ matter obtained with 6 versions of the density dependent
NN interaction (\ref{eq2})-(\ref{eq4}) is shown in Fig.~\ref{f6}. Like the Love 
number and tidal deformability of NS shown in Figs.~\ref{f4} and \ref{f5}, the 
$M$-$R$ results obtained with 6 versions of the interaction comply well with 
the GW170817 constraint for NS with mass $M=1.4~M_\odot$ \cite{Abbott18} 
(the colored contours in Fig.~\ref{f6}). However, only the EOS's obtained with 
the CDM3Y6, CDM3Y8 and BDM3Y1 interactions give the maximum mass $M$ close 
to the observed lower mass limit of PSR J$0348+0432$ and PSR J$0740+6620$. 
It is interesting to note that the $M$-$R$ curves obtained with 6 interactions
seem to coincide at $M\approx 0.55~M_\odot$ and $R\approx 11.75$ km, and such 
a $M$-$R$ point results from the behavior of the pressure of NS matter at the 
saturation density $n_0$ discussed in Fig.~\ref{f3}.   

Although the impact of the nuclear incompressibility on the EOS of NM is well 
established \cite{Schu96,Kho07r}, much more efforts are made nowadays to explore 
the impact of the nuclear symmetry energy on the macroscopic properties of NS, 
like the sensitivity of the mass and radius of NS to the strength and slope 
of $E_{\rm sym}(n_{\rm b})$ at high baryon densities \cite{Loan11,Xie19}. 
The results discussed in the present work remind us again of the important 
effect by the nuclear incompressibility $K$ in the nuclear mean-field description 
of the EOS of NS matter. The results shown in Fig.~\ref{f6} suggest that a stiffer 
EOS of NS matter associated with $K\approx 250-270$ MeV seems to comply better 
with the constraints implied by the gravitational wave observations of the NS 
merger and observed masses of the heaviest neutron stars. This result is complementary 
to the recent joint analysis of the NICER and LIGO/Virgo data \cite{Raa20}, which 
prefers a stiff EOS associated with the masses of the heaviest pulsars observed so far.   

It should be noted that the fast-rotating millisecond pulsar J$0740+6620$ might 
be associated with a strong magnetic field, so that the EOS of NS matter could also 
be affected by the spin polarization of baryons \cite{Lat07}. In particular, the maximum 
mass of magnetar might increase by about $0.1~M_\odot$ if $60\sim 70\%$ of baryons 
in NS matter are spin-polarized \cite{Tan20}. In such a scenario, the heavy mass 
($M\approx 2.14^{+0.20}_{-0.18}~M_\odot$) observed for the millisecond 
PSR J$0740+6620$ \cite{Cro20} could have imprints from both the spin polarization 
of baryons and incompressibility of NS matter. This should be the subject of further 
mean-field studies.       

\section*{Summary}
The EOS of asymmetric NM has been studied based on the nuclear mean-field 
potentials given by the nonrelativistic HF calculation using different versions 
of the density dependent M3Y interaction that are associated with the nuclear 
incompressibility $K\approx 217-270$ MeV at the saturation density $n_0$. From these 
scenarios of the EOS, only the CDM3Y6 and CDM3Y8 versions give the total energy 
of symmetric NM and pure neutron matter at high baryon densities close to the results 
of the ab-initio calculations of NM \cite{APR,MMC}.   

The HF mean-field potential is used to generate the EOS of $\beta$-stable 
np$e\mu$ matter, which is then compared with the empirical EOS 
(in terms of the total pressure of NS matter) deduced from the Bayesian 
analysis of the GW170817 data at the 90\% confidence level \cite{Abbott18}.
We found that only the pressure of NS matter given by the EOS's obtained with 
the CDM3Y6, CDM3Y8, and BDM3Y1 interactions (associated with $K\approx 252$, 
257, and 270 MeV, respectively) agrees with the empirical pressure at high 
baryon densities, up to around $6n_0$. 

The EOS of $\beta$-stable np$e\mu$ matter is also used to determine 
the tidal deformability $\Lambda$, mass $M$ and radius $R$ of NS using 
Eqs.~(\ref{eq19})-(\ref{eq22}) derived from the General Relativity. The
obtained results show clearly the impact by the nuclear incompressibility $K$ 
on the mean-field description of NS matter. While $\Lambda_{1.4}$ and $R_{1.4}$ 
given by all versions of the density dependent NN interaction are well inside 
the corresponding empirical ranges imposed for NS with mass $M=1.4~M_\odot$ 
by the Bayesian analysis of the tidal wave signals from the NS merger GW170817 
\cite{Abbott18}, only the EOS associated with $K\approx 250-270$ MeV gives 
the maximum gravitational mass of NS comparable to that deduced for the 
second PSR J$0348+0432$ and millisecond PSR J$0740+6620$, the heaviest 
neutron stars observed so far.  

\begin{acknowledgement}
\textbf{Acknowledgement} 
We thank N. H. Phuc for his help in revising the isovector density dependence
of the M3Y interaction. The present research was supported, in part, by the 
National Foundation for Science and Technology Development of Vietnam 
(NAFOSTED Project No. 103.04-2017.317).
\end{acknowledgement}

\end{document}